\documentclass[12pt]{article}

\usepackage{scicite}

\usepackage{times}
\usepackage{graphicx}%
\usepackage{multirow}%
\usepackage{amsmath,amssymb,amsfonts}%
\usepackage{amsthm}%
\usepackage{mathrsfs}%
\usepackage[title]{appendix}%
\usepackage{xcolor}%
\usepackage{textcomp}%
\usepackage{manyfoot}%
\usepackage{booktabs}%
\usepackage{algorithm}%
\usepackage{algorithmicx}%
\usepackage{algpseudocode}%
\usepackage{listings}%

\newenvironment{sciabstract}{%
\begin{quote} \bf}
{\end{quote}}

\title{Coherent Thermal Emission from Large-Scale Suspended Nanomechanical  Membranes } 

\author
{Mitradeep Sarkar,$^{1}$ Rajashree Haldankar,$^{1}$ Julien Legendre,$^{1}$ Gloria Davidova,$^{1}$ \\
Adrian Bachtold,$^{1}$ Georgia T. Papadakis,$^{1\ast}$\\
\normalsize{$^{1}$ICFO-Institut de Ciencies Fotoniques, The Barcelona Institute of Science and Technology,}\\
\normalsize{Avinguda Carl Friedrich Gauss, 3, 08860 Castelldefels, Barcelona, Spain}\\
\normalsize{$^\ast$To whom correspondence should be addressed; E-mail:  georgia.papadakis@icfo.eu.}
}

\date{}
\begin{document} 
\baselineskip24pt
\maketitle 
\begin{sciabstract}
Thermal radiation is an abundant form of incoherent light. Generating coherent infrared light through incandescence promises a cheap alternative to the costly and epitaxially complex quantum cascade laser, however it remains a fundamental challenge. Previous approaches leveraged the spatial coherence of polaritonic excitations that occur in the thermal near-field, by diffracting them into the far-field zone via patterned micro- or nano-scatterers. This approach requires high-resolution lithography, is difficult to scale-up, and yields limited outcoupled radiation due to the intrinsically polarized nature of polaritons. We overcome these limitations and report coherent thermal emission through simple wave interference. We show that unpatterned, millimeter-scale, suspended nanomechanical membranes of SiC operate for both linear polarizations and exhibit antenna-like directionality without relying on the excitation of near-field polaritons. The ability to generate polarization-insensitive, narrowband and spatially coherent incandescent light without lithography at large scales paves the way towards democratizing thermal infrared technologies.
\end{sciabstract}


\par{Incandescence is the emission of thermal radiation by black or grey bodies, which peaks at mid-infrared (IR) frequencies for near-room temperatures, as predicted by Planck's law \cite{Planck1901}. Detecting deviations from the black body spectrum in the thermal emissivity of materials is central to applications in sensing, spectroscopy, and materials' identification, while tailoring this emissivity by design enables contactless temperature regulation, thermal camouflage, IR imaging, and thermophotovoltaic energy conversion \cite{Baranov2019}. Many of these applications require a coherent IR light source \cite{Hlavatsch2022}. Examples include thermophotovoltaic systems operating at maximal efficiency \cite{Papadakis2019,Sakakibara2019}, detection and identification of molecular fingerprints \cite{Rodrigo2015,Oh2021biosensors,Zhang2024}, IR holography \cite{Ravaro2014}, image encryption via physical tags \cite{Audhkhasi2023} and multispectral imaging \cite{Tittl2015} that require narrowband and directional response of individual pixels. Nonetheless, generating coherent light at mid-IR frequencies presents fundamental challenges; light emitting diodes (LEDs) suffer from strong non-radiative losses \cite{Higashitarumizu2023} and are thus alien to the spectral range above 5 $\mathrm{\mu m}$ \cite{Hlavatsch2022}, while quantum cascade lasers (QCL) rely on expensive epitaxial techniques. Due to the ubiquitous nature of thermal radiation, IR light emission through incandescence promises a cheap alternative to the QCL, however conventional incandescent sources and filaments, like globars, emit radiation with similar characteristics to that of a black body. Thereby, they generate incoherent light that significantly reduces their efficiency due to parasitic emission towards unwanted frequencies and directions.}

\par{Generating coherent incandescent light requires confining the spectrally broad and spatially diffuse characteristics of black body radiation into a narrow spectral and angular range. This becomes possible by tailoring the thermal emissivity via nanophotonic design \cite{Baranov2019,Shen2016,Qu2020,Greffet2002}. To reduce the spectral bandwidth of far-field thermal radiation at mid-IR frequencies, one can harness the long lifetimes of phonon-induced resonances in crystalline polar materials like silicon carbide (SiC) \cite{Caldwell2015, Basov2016, Foteinopoulou2019}, or utilize the quantum confinement in low-dimensional materials such as carbon nanotubes \cite{Mueller2010} and atomically thin semiconductors \cite{Dobusch2017}. It is also possible to leverage wave interference to induce narrowband emission by employing the concept of a Salisbury screen \cite{salisbury_1952}. In a Salisbury screen, absorption in a lossy thin-film is resonantly enhanced through interference; the absorbing layer is separated from a back-side reflector via a quarter-wavelength-thick dielectric spacer that enables constructive interference \cite{ergoktassciencetopo, Fang2018, Jung2014}. By Kirchhoff's law of thermal radiation \cite{Kirchhoff1860}, a Salisbury absorber can also serve as an emitter, in which case the lossy thin-film serves as the emitting layer. As shown by Ergoktas recently \cite{ergoktassciencetopo}, precise tuning of the thickness of the emitting layer yields localized emission, which however remains spatially diffuse.}

\par{To narrow the spatially diffuse characteristics of black body radiation, various approaches have been developed since the instrumental work by Greffet \textit{et al.} \cite{Greffet2002}. The coherent nature of thermally excited surface polaritons in the near-field, in other words at microscopic distances from a surface, can yield antenna-like directional emission in the far-field zone \cite{PhysRevLett.82.1660}. The concept was first demonstrated on a SiC surface that supports surface-phonon polaritons. Via a lithographically patterned grating, these polaritons diffracted into far-field propagating electromagnetic modes towards specific angles. Similar demonstrations that rely on surface-plasmon polaritons have also been reported \cite{plasmonicmetagreffet,Han:10}. However, surface plasmon- or phonon-polaritons occur only in transverse magnetic (TM) polarization \cite{Papadakis.PhysRevMaterials}, hence, concepts that rely on their excitation are generally polarization-specific. In addition, they require coupling between the thermal near-field and far-field zones. In the landscape of thermal emissivity engineering, enabling this coupling requires diffractive elements such as gratings and nanoantennas \cite{Lu2021phononcoupling,Yu2023,Baranov2019,Shen2016,Qu2020,Greffet2002,Han:10}, for which micro- or nano-patterning is required. This demonstrates the need for high-resolution lithography that hinders large-scale adoption of coherent thermal sources. Additionally, relying on the excitation of TM-polarized polaritons limits the amount of outcoupled radiation and thereby reduces by half the luminocity of a potential thermal source.} 


\par{To realize technologically relevant mid-IR thermal sources, it is critical that spatial and spectral coherence do not come at the cost of complicated lithographic steps. At the same time, to maximize the brightness or luminocity of a source, it is critical that the thermal emission is not polarization-specific. Recently, directional thermal emission with planar, pattern-free structures has been demonstrated in the context of the epsilon-near-zero response of materials \cite{Jin2021, Hwang2023}, where a Berreman mode occurs \cite{Vassant:12}. In both \cite{Jin2021} and \cite{Hwang2023}, however, the reported emissivity remained spectrally incoherent, occurred only for TM polarization, and the reported directionality was considerably inferior to nano-structured surfaces \cite{Greffet2002,Han:10} . Alternatively, directional emission is possible with periodic arrangements of alternating lossy bilayers in a photonic crystal \cite{Qu2020, Liu2019}. However, for strong directionality, more than ten bilayers are required to mimic the response of a bulk photonic crystal, making their realization impractical and expensive.}

\par{Here, we introduce a radically different approach to confine thermal radiation into an ultra-narrow spectral and spatial range that does not rely on the near-field coherence of polaritons \cite{PhysRevLett.82.1660,Greffet2002,Han:10}, neither on collective excitations in subwavelength meta-architectures \cite{plasmonicmetagreffet,Lu2021phononcoupling,Yu2023,Baranov2019,Shen2016,Qu2020,Tittl2015}, nor on photonic crystal effects \cite{Qu2020, Liu2019, Qian2017,Hlavatsch2022}. We experimentally demonstrate large-scale (lateral dimensions $\sim2$ mm), lithography-free thermal emission arising from suspended nanomechanical membranes of SiC, with thickness $200$ nm. In analogy to the work by Greffet \textit{et al.} \cite{Greffet2002}, we utilize SiC as the thermally emitting material, however, without patterning it. The angular selectivity becomes possible by integrating the membranes into a modified Salisbury screen. Although both a single slab of SiC and a SiC-based conventional Salisbury screen yield diffuse emission \cite{ergoktassciencetopo}, by modifying the characteristic dimensions of a Salisbury screen we enable previously unreported antenna-like directionality. This directionality is the result of wave interference, thereby, the resulting thermal emission is polarization-insensitive. At a central wavelength of $13.2$ $\mathrm{\mu}$m, we measure a narrow spectral emission bandwidth of $0.7$ $\mathrm{\mu}$m and obtain emission lobes with an angular spread of $12.9$ degrees. The angular spread is measured via direct thermal emission measurements, and is comparable to the values reported in the case of gratings \cite{Greffet2002} and 1D photonic crystals \cite{Liu2019} when conducting similar thermal emissivity characterization above room temperature.}

\par{We note that, although Salisbury screens are trivial to realize at other spectral ranges, their experimental implementation at mid-IR frequencies remains a challenge. In particular, a key constituent of a Salisbury screen is a transparent dielectric spacer, however at mid-IR frequencies most materials resonantly absorb due to crystal lattice vibrations \cite{Caldwell2015}, thereby there is a scarcity of lossless and dispersionless mid-IR transparent media. Here, we show an alternative and versatile approach to building Salisbury screens at mid-IR frequencies by considering an air gap as the dielectric spacer, for which the SiC nanomembranes are mechanically suspended.}

\section*{Design of coherent thermal source}

\par
\begin{figure}[h]
\centering
\includegraphics[width=0.9\textwidth]{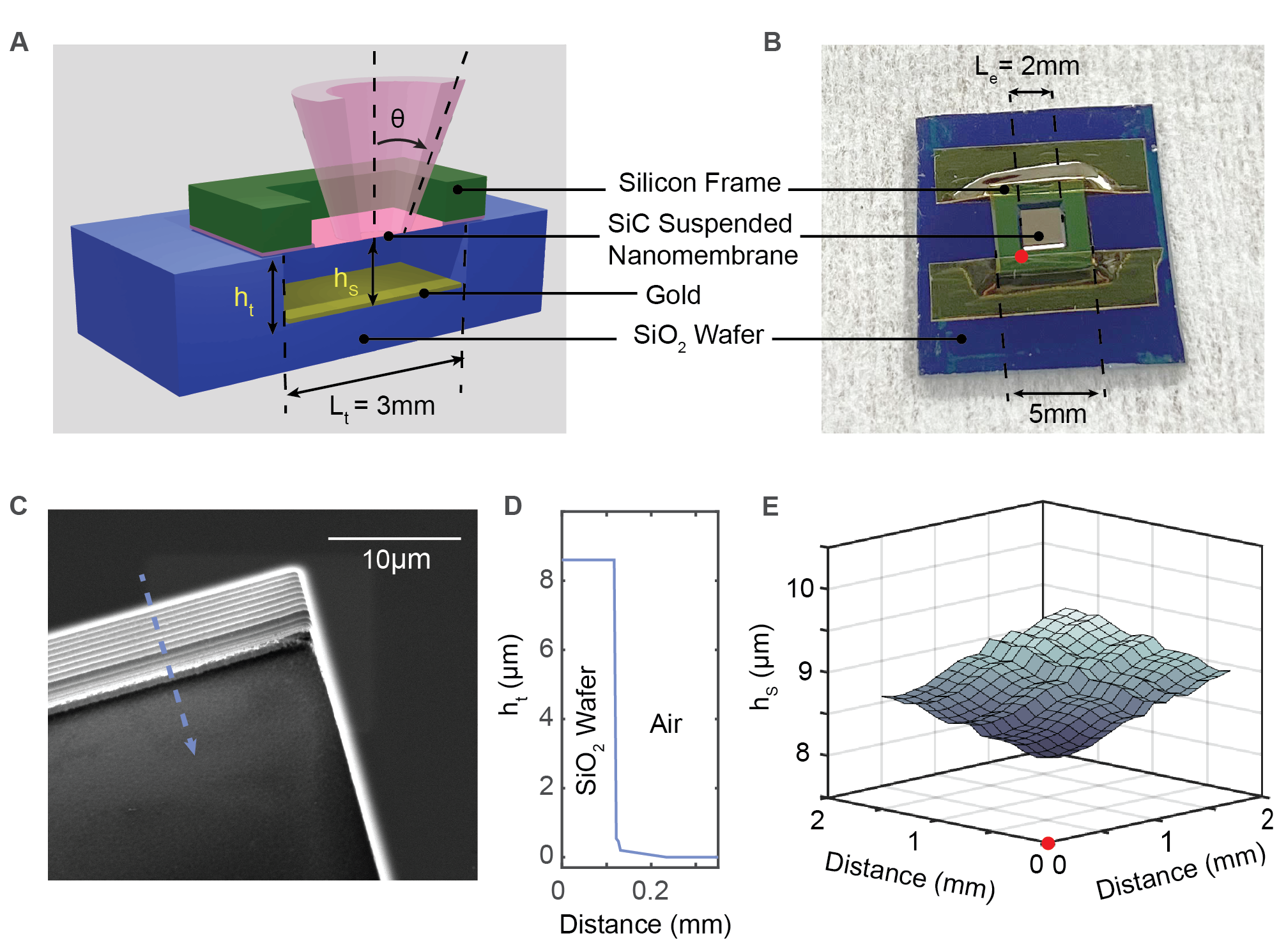}
\caption{\textbf{Fabrication and geometrical specifications of the emitter.} \textbf{(A)} Schematic of the thermal source: a modified Salisbury screen with an air-spacer layer and SiC emitting layer. \textbf{(B)} A false-color photograph image of the fabricated device. \textbf{(C)} SEM image of one of the corners of the etched trench. \textbf{(D)} Height profile ($h_\mathrm{t}$) of the etched trench measured by profilometry. The position where this measurement was conducted on the sample is shown in Fig. \ref{fig1} (C) with the dashed blue line. \textbf{(E)} Spacer thickness ($h_\mathrm{s}$) measured by FTIR micro-spectroscopy. The reference point for these measurements is shown with a red dot here and in panel (B).}\label{fig1}
\end{figure}

{A Salisbury screen provides near-unity thermal emissivity on-resonance from a planar three-layered heterostructure. It consists of an infinitesimally thin lossy emitting layer and a $\lambda/4$-thick spacer layer on a back-side reflector, where $\lambda$ is the wavelength of light inside the spacer. The conventional Salisbury configuration yields narrowband emission but the emitted light is spatially diffuse \cite{salisbury_1952,ergoktassciencetopo}. Without loss of generality, here we consider that the dielectric spacer air ($n_\mathrm{s}=1$).

\par{We recently demonstrated that, in order to achieve maximal thermal emission in the three-layered configuration of a Salisbury screen, the following condition should be satisfied \cite{Sarkar2024}}:
\begin{equation}\label{eq:phase matching}
	\frac{4\pi}{\lambda} h_\mathrm{s}\cos{\theta}+2\Psi_\mathrm{e}=(2l-1)\pi,
\end{equation}
\noindent{where $h_\mathrm{s}$ is the thickness of the dielectric spacer, and $\theta$ is the central angle of emission (see Fig. \ref{fig1} (A)). The parameter $\Psi_\mathrm{e}$ represents the phase accumulated upon single pass within the lossy emitter, and depends on both the refractive index ($n_\mathrm{e}$) and thickness ($h_\mathrm{e}$) of this layer. In a conventional Salisbury screen, the phase accumulated inside the lossy emitter vanishes due to its small thickness ($\Psi_\mathrm{e} \xrightarrow[h_\mathrm{e} \to 0] {} 0$), for which Eq. \ref{eq:phase matching} yields $h_\mathrm{s}=\lambda/4$. By contrast, directional thermal emission is achieved when $\Psi_\mathrm{e}\approx\pi/2$ , for which Eq. \ref{eq:phase matching} yields $h_\mathrm{s}=\lambda/(2\cos\theta)$. The large ($\pi/2$) phase accumulation inside the emitter is observed as strong light confinement in \cite{Sarkar2024}, making the device operate like an optical cavity, thereby assuring spatially coherent emission (see Supplementary Information). To achieve $\Psi_\mathrm{e}\approx\pi/2$, the refractive index of the lossy emitting layer should be much larger than unity, and its thickness should no longer be small as in the original Salisbury configuration \cite{salisbury_1952}. Considering $h_\mathrm{e}=\lambda/(4 \lvert n_\mathrm{e}\rvert)$ warrants sufficient phase accumulation while Eq. \ref{eq:phase matching} remains satisfied.}

We consider SiC for the lossy emitting layer, since it supports an ultra-high refractive index in its crystalline form at frequencies near its Reststrahlen band, corresponding to wavelengths in the range $12.6$ $\mathrm{\mu}$m - $13.5$ $\mathrm{\mu}$m. Importantly, this is the spectral range where the black body spectrum is thermal radiation is maximal for near-room temperatures \cite{Caldwell2015, Foteinopoulou2019}. For wavelengths in this range, the SiC layer should have a thickness of $h_\mathrm{e}=\lambda/(4 \lvert n_\mathrm{e}\rvert)\approx 200$ nm.}

\par{Central to realizing a Salisbury screen at mid-IR frequencies is to identify appropriate non-absorbing and dispersionless transparent materials that can serve as the dielectric spacer layer \cite{symmons2021field} as well as appropriate methods to grow or deposit them. The targeted thickness for the spacer layer is in the range of few-microns, therefore conventional deposition methods, such as thermal and electron beam evaporation or reactive sputtering, cannot be used due to challenges in film uniformity, adhesion, buckling, peel-off, and related effects \cite{dielectricfilmdipos}. In addition, most dielectrics exhibit strong polar resonances in the mid-IR range. To circumvent these challenges, we utilize an air gap as a dielectric spacer (see schematic in Fig. \ref{fig1} (A)). Due to its low-refractive index, air is an ideal spacer for directional emission \cite{Sarkar2024}. By aiming to achieve directional thermal emission at an angle of approximately $\theta_\mathrm{0}=40$ degrees and at wavelengths near the Reststrahlen band of SiC, the height of the air gap should be $h_\mathrm{s}=\lambda/(2\cos\theta_\mathrm{0})\approx8.5\mathrm{\mu m}$.}

\par{To practically realize this air gap, we etch a trench of height $h_\mathrm{t}=8.5$ $\mu$m and lateral dimensions $L_\mathrm{t}$ $\times$ $L_\mathrm{t}$, where $L_\mathrm{t}=3$ mm, into a silicon dioxide wafer ($\mathrm{SiO_2}$) via reactive ion etching. A corner of the trench is shown via a scanning electron microscope (SEM) image in Fig. \ref{fig1} (C), and its height is measured via profilometry as shown in Fig. \ref{fig1} (D). A thin ($150$ nm) layer of gold is deposited at the bottom of the trench by thermal evaporation. This gold layer serves as the back-side reflector of the Salisbury screen. We place on top of the trench the $200$ nm-thick layer of SiC, which is in the form of a commercially available nanomechanical suspended membrane with lateral dimensions $L_\mathrm{e}$ $\times$ $L_\mathrm{e}$, where $L_\mathrm{e}=2$ mm, purchased from Norcada Inc. The membrane is commercialized with a $400$ $\mu$m-thick silicon frame that is shown with the green color in Fig. \ref{fig1} (A). A photograph of the complete device is shown in Fig. \ref{fig1} (B).}

\par{The lateral dimensions of the etched trench are intentionally selected to be slightly larger than those of the SiC nanomembrane to avoid direct contact of the trench edges with the nanomembrane in order to reduce mechanical stress  \cite{Chen2016, Jung2014}. The flatness of the nanomembrane upon its integration onto the trench is evaluated by scanning the whole area of the membrane while conducting micro-Fourier Transform Infrared Spectroscopy (FTIR). In particular, we measured the thickness of the spacer layer, $h_\mathrm{s}$, at each scanned position, by tracking the reflectance minimum at normal incidence, which occurs when Eq. \ref{eq:phase matching} is satisfied for $\theta=0$. The measured $h_\mathrm{s}$ is shown in Fig. \ref{fig1} (E). As shown in Fig. \ref{fig1} (E), $h_\mathrm{s}$ varies from $8.5$ $\mu$m to $9.2$ $\mu$m, a variation that does not considerably affect the degree of directionality as shown in the following results. A small tilt of the nanomembrane towards one of the corners of the wedge is seen from Fig. \ref{fig1} (E). Despite this tilt, the suspended nanomembrane exhibits no bending or buckling across a surface area of $2$ mm $\times$ $2$ mm despite its high length-to-thickness aspect ratio $h_\mathrm{e}/L_\mathrm{e}=10^{-4}$, due to the high stress homogeneity in the suspended nanomembrane \cite{youngSiC}. We note that, in extracting the thickness of the spacer, we utilized experimentally measured values of the refractive index of SiC (see Supplementary Information Section 2), which agree with those reported in the literature for crystalline SiC \cite{Spitzer1959}. For additional details on the experimental realization, device imaging and characterization, see Supplementary Information.}

\section*{Experimental characterization of directional absorptivity and thermal emission}

\begin{figure}[h]
\centering
\includegraphics[width=1\textwidth]{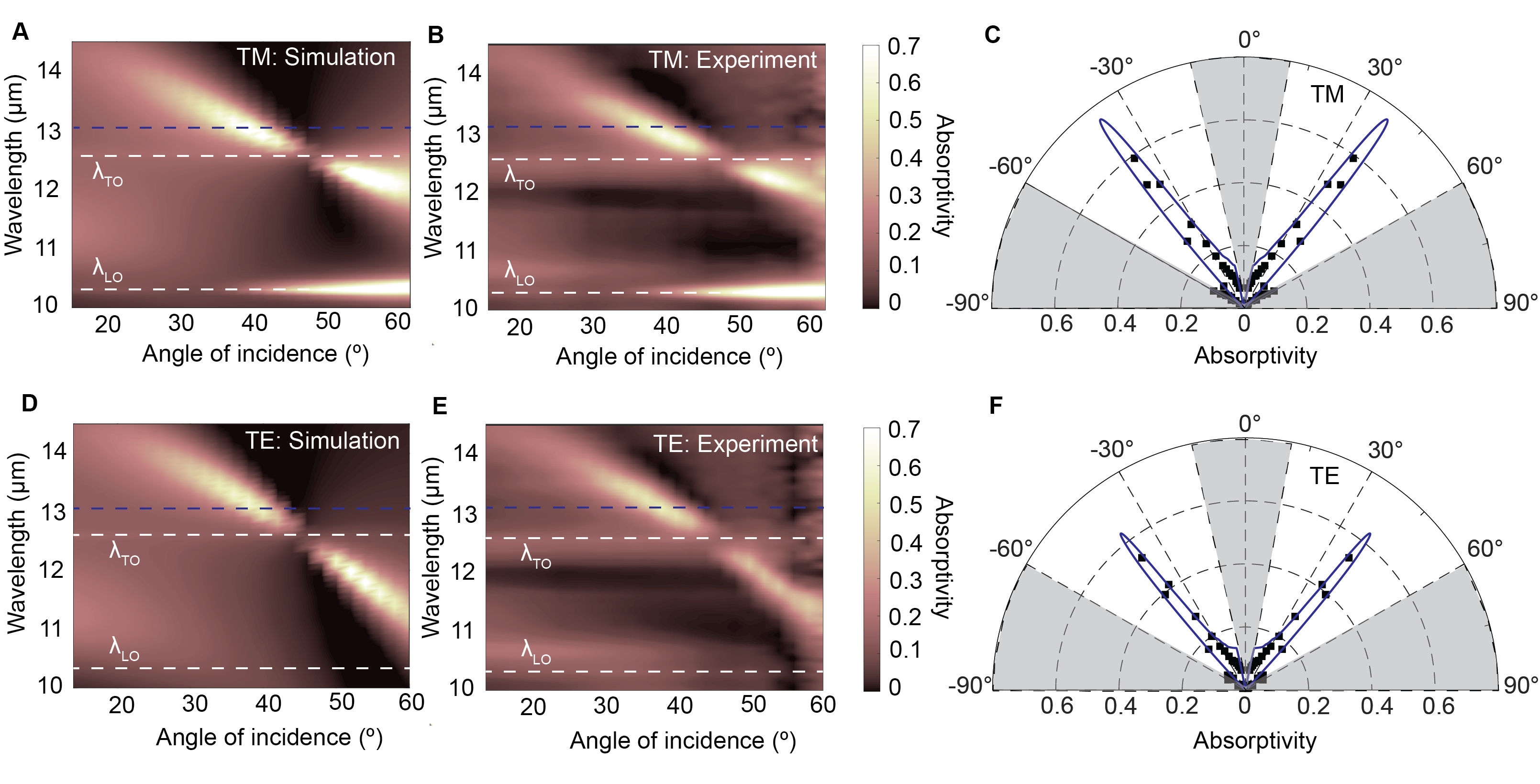}
\caption{ \textbf{Angle-dependent absorptivity spectra.} \textbf{(A)} Calculated absorptivity ($\mathcal{A}=1-R$) as a function of wavelength, $\lambda$, and the angle of incidence, $\theta$, for TM polarized light, averaged over the range of $h_\mathrm{s}$ indicated in Fig. \ref{fig1} (E). \textbf{(B)} Corresponding experimental measurement of absorptivity for TM polarization. \textbf{(C)} Polar plot of the absorptivity for TM polarization where solid blue lines correspond to theoretical predictions solid black dots correspond to experimental measurements, for the wavelength of $\lambda=13.2$ $\mu$m. \textbf{(D)} Calculated absorptivity for TE polarized light and \textbf{(E)} corresponding experimental measurement. \textbf{(F)} Polar plot of $\mathcal{A}$ for TE polarization at $\lambda=13.2$ $\mu$m. In panels (A), (B), (D), (E), the solid horizontal lines correspond to the TO and LO phonon resonances of SiC. The dashed blue horizontal line corresponds to the wavelength for which the polar plots in panels (C), (F) are shown.}\label{fig2}
\end{figure}

\par{In the absence of magnetic effects, Kirchhoff's law of thermal radiation \cite{Kirchhoff1860} imposes an equality between the thermal emissivity, $\mathcal{E}$, and the absorptivity, $\mathcal{A}$, per frequency, polarization, and direction. In opaque substances with vanishing transmission, the absorptivity equals $\mathcal{A}=1-R$, where $R$ is the reflectivity. In the experimental characterization of the thermal source, we conducted both angle-dependent absorptivity and emissivity measurements. The absorptivity measurements are carried out by probing the reflectivity for both linear polarizations: transverse magnetic (TM) and transverse electric (TE), using a manual variable-angle reflection stage coupled to an FTIR. The measured $\mathcal{A}$, shown in Figs. \ref{fig2} (B), (E) for TM and TE polarization, respectively, were normalized to a reference gold mirror. In Figs. \ref{fig2} (A), (D), we show the corresponding calculations that were carried out using the transfer matrix method \cite{mtenders_2024_10654406}, averaged over the range of spacer layer thicknesses that were experimentally measured as shown in Fig. \ref{fig1} (E)  (see Supplementary Information Section 5).}

\par{As shown in Fig. \ref{fig2} (A), (B) and (D), (E), the calculated and measured absorptivities, respectively, are in good agreement. In these panels, the wavelengths corresponding to the transverse optical (TO) and longitudinal (LO) phonons of SiC are marked with $\lambda_\mathrm{TO}$ and $\lambda_\mathrm{LO}$, respectively (see dashed horizontal lines). As expected from theory \cite{Sarkar2024}, for both linear polarization, absorptivity maxima occur near $\lambda_\mathrm{TO}$. By contrast, the additional absorptivity maximum near $\lambda_\mathrm{LO}$ corresponds to the excitation of a surface phonon polariton that occurs only for TM polarization (panels (A) and (B)). The polarization-independent absorptivity maxima near $\lambda_\mathrm{TO}$ shift as a function of $\theta$, similar to the case of a diffraction grating \cite{Greffet2002}. In Figs. \ref{fig2} (C), (F) we show the polar plots of the absorptivity at the wavelength of $\lambda=13.2$ $\mu$m, for TM and TE polarization, respectively (this wavelength is marked with a dashed blue line in panels (A), (B), (D), (E) of the same figure). As seen, the solid blue lines, representing the theoretical calculation, match the experimental measurements, shown with solid black dots. The angular width of the lobes for TM polarization is measured to be $7.4$ degrees, whereas the theoretical calculation yields $7.7$ degrees. For TE polarization, the measured as well as the theoretically calculated angular width is $6.5$ degrees. From Fig. \ref{fig2}, the thermal source preferentially absorbs mid-IR light at specific wavelengths and incident angles with directionality comparable to nanostructured surfaces \cite{Greffet2002}.}  
 
\begin{figure}[h]
\centering
\includegraphics[width=0.9\textwidth]{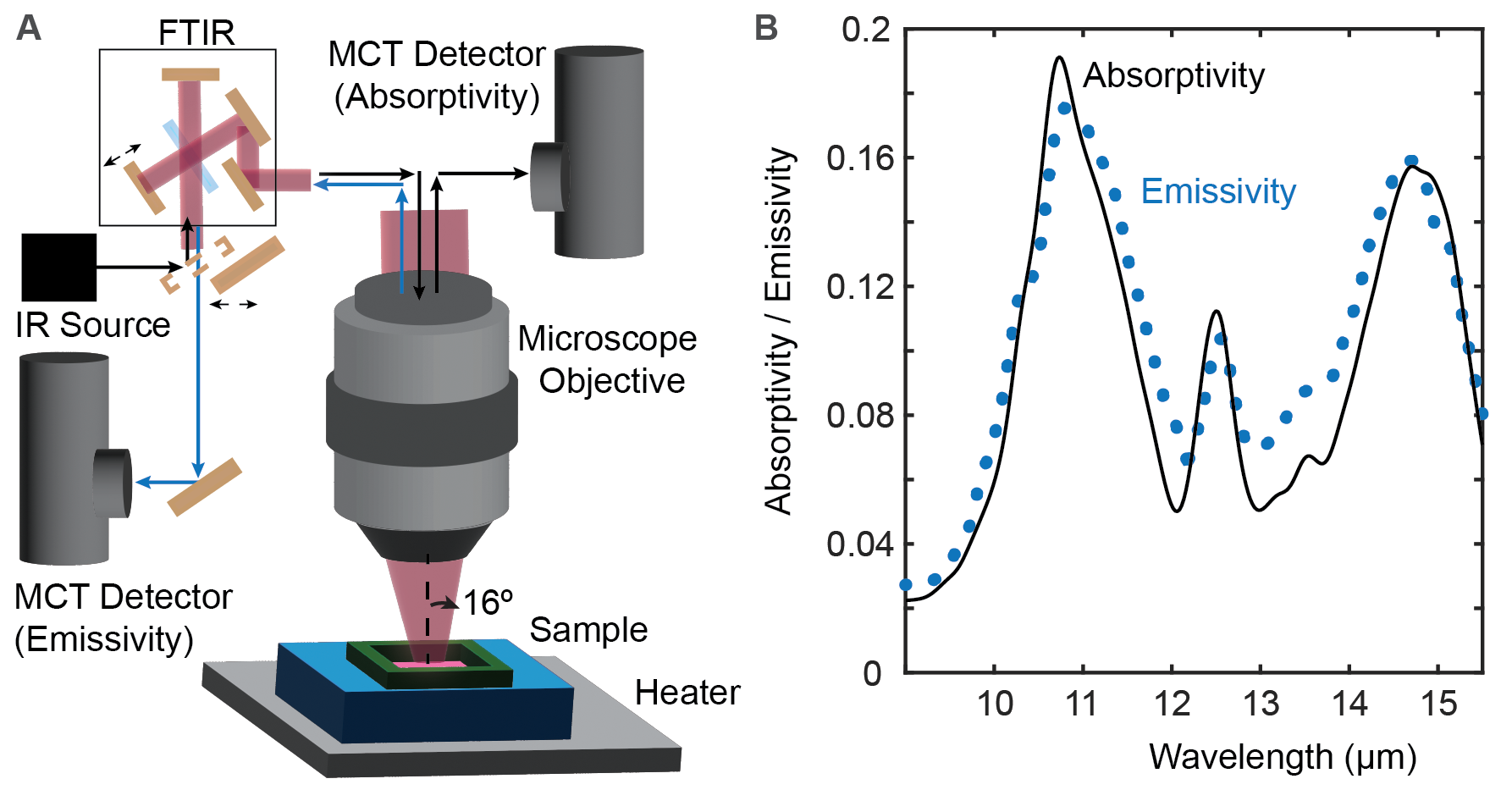}
\caption{\textbf{Simultaneous absorptivity ($\mathcal{A}$) and emissivity ($\mathcal{E}$) measurement at normal incidence.} \textbf{(A)} Schematic of the FTIR micro-spectroscopy setup. We show the beam path for $\mathcal{A}$ and $\mathcal{E}$ measurements by black and blue arrows respectively. \textbf{(B)} The spectra of the measured $\mathcal{A}$ (solid black curve) and $\mathcal{E}$ (blue dots) from the same spot on the SiC nanomembrane.}\label{fig3}
\end{figure}

\par{Next, we carry out direct thermal emissivity measurements by heating the sample and probing its thermal emission. To probe thermal emissivity and absorptivity simultaneously, these measurements are first conducted at normal incidence, $\theta=0$, from a spot of $100\mathrm{\mu m}\times 100\mathrm{\mu m}$, using FTIR micro-spectroscopy with a Cassegrain objective that has an aperture angle of $16$ degrees. The schematic of the experimental setup is shown in Fig. \ref{fig3} (A). As shown, two separate MCT detectors are used for these measurements; in the beam path for the absorptivity measurements, the sample is placed between the FTIR and the detector. By contrast, in the emissivity measurements the blackbody source of the FTIR is replaced by the sample itself, thus changing the beam path.}

\par{To extract the thermal emissivity we adopt the method outlined by Xiao \textit{et al.} in \cite{Xiao2019}, by first measuring the emission signal $S(\lambda,T)=m(\lambda)\left[\mathcal{E}(\lambda)I_\mathrm{BB}(\lambda,T)+B(\lambda)\right]$, 
where $m(\lambda)$ is the spectral response of the measuring system, $I_\mathrm{BB}(\lambda,T)$ is the black body radiation given by Planck's law, and $B(\lambda)$ is the background emission at room temperature. We measure the emission signal from a carbon tape, which serves as the reference, $S_\mathrm{ref}$, and that of the sample, $S_\mathrm{sample}$, at two temperatures: $T_\mathrm{1}=423K$ and $T_\mathrm{2}=393K$, using a thermal stage (Linkam Inc.). The sample's emissivity, $\mathcal{E}(\lambda)$, is calculated as:
\begin{equation}\label{eq:emissivity}
	\mathcal{E}(\lambda)=\mathcal{E}_\mathrm{ref}\left[\frac{S_\mathrm{sample}(\lambda,T_\mathrm{1})-S_\mathrm{sample}(\lambda,T_\mathrm{2}) }{S_\mathrm{ref}(\lambda,T_\mathrm{1})-S_\mathrm{ref}(\lambda,T_\mathrm{2})}\right],
\end{equation}
where $\mathcal{E}_\mathrm{ref}$ is the emissivity of the carbon tape, which was evaluated by measuring the reflectance from the tape (see Supplementary Information Section 6).} 

\par{The simultaneous measurements of absorptivity and emissivity at normal incidence are shown in Fig. \ref{fig3} (B). Since thermally emitted light is intrinsically unpolarized, the incident light for the absorptivity measurements was also unpolarized for these measurements, for a direct comparison between $\mathcal{E}$ and $\mathcal{A}$. As shown, the two measurements agree, as expected, with small deviations. These deviations are expected since, at the measured temperatures that are not considerably higher than room-temperature, the emitted signal from the sample is comparable in magnitude to the background thermal emission arising from the instrument itself (see Supplementary Information). Thereby, there is a higher uncertainty associated with measuring $\mathcal{E}$ than measuring $\mathcal{A}$. There exist three peaks in both $\mathcal{A}$ and $\mathcal{E}$ spectra. The two peaks at the shorter wavelengths result from the intrinsic properties of SiC, in particular its vanishing permittivity at the wavelengths of $\lambda_\mathrm{LO}$ and $\lambda_\mathrm{TO}$ (see Fig. \ref{fig2} and Supplementary Information). These are independent from the angle of observation. The peak near $\lambda=14.7\mathrm{\mu m}$ corresponds to a Fabry-Perot mode that also satisfies Eq. \ref{eq:phase matching} for $\theta=0$. This peak, as shown in Fig. \ref{fig2}, depends strongly on the observation angle.}

\begin{figure}[ht!]
\centering
\includegraphics[width=0.9\textwidth]{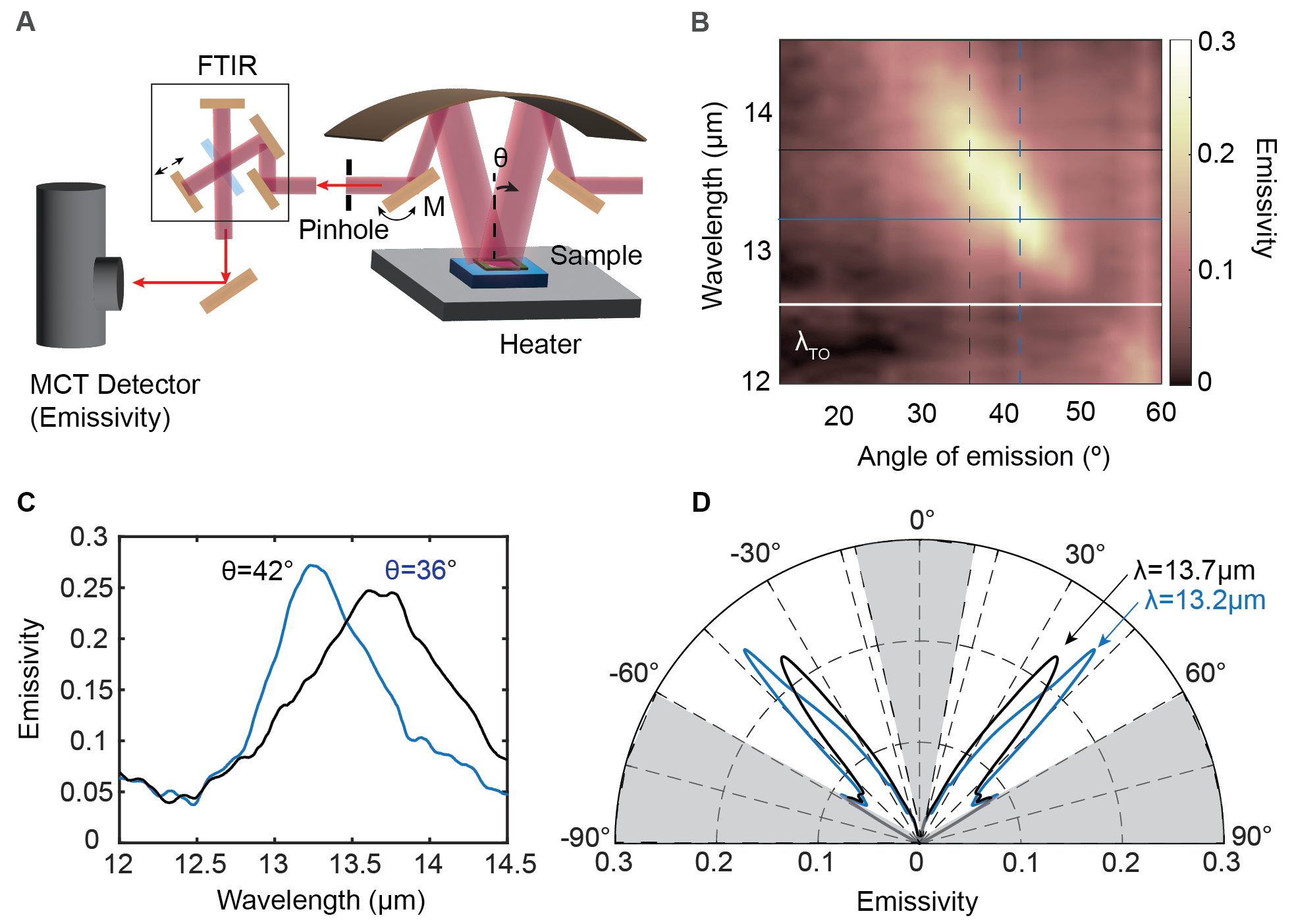}
\caption{\textbf{Angle-dependent emissivity spectra.} \textbf{(A)} Schematic of the experimental setup to measure emissivity ($\mathcal{E}$) as a function of wavelength, $\lambda$, and angle of emission, $\theta$. The observation angle $\theta$ was changed manually using a rotating mirror noted as \textbf{M}. \textbf{(B)} Experimental measurement of $\mathcal{E}$ for unpolarized light. The solid horizontal lines correspond to the TO phonon resonances of SiC (white) as well as the two wavelengths plotted in the polar plots of panel (D). The dashed vertical lines correspond to the angles for which spectra are shown in panel (C). \textbf{(C)} Emissivity spectra at $\theta=36$ and $42$ degrees. \textbf{(D)} Polar plot of the measured emissivity at $\lambda=13.2\mathrm{\mu m}$ and $\lambda=13.7\mathrm{\mu m}$.}\label{fig4}
\end{figure}

\par{Next, we conduct angle-dependant measurements of the thermal emissivity using the same manual variable angle stage coupled to the same FTIR that was used for absorption measurements. The schematic of the experimental setup is shown in Fig. \ref{fig4} (A). Similar to the measurements at normal incidence (Fig. \ref{fig3}), for each angle of observation, we obtain the emission spectra for the sample and the carbon tape at $T_\mathrm{1}=423K$ and $T_\mathrm{2}=393K$. From these spectra, the angle-dependent thermal emissivity is derived (see Eq. \ref{eq:emissivity} and Supplementary Information) and presented in Fig. \ref{fig4} (B). As expected, it closely follows the dispersion of the absorptivity shown in Figs. \ref{fig2} (B), (E), with a major difference being that the emissivity measurements are conducted for unpolarized light whereas the absorptivity measurements in Fig. \ref{fig2} were conducted while polarizing the incident beam.}

\par{To suppress the influence of the background thermal emission from the FTIR instrument itself, which lowers the signal-to-noise ratio that is detected, we used a pinhole to limit the area of the sample from which emission is detected. Although such a pinhole was also used in the absoprtivity measurements, the aperture size of the pinhole was different in the two measurements. The pinhole's diameter in the absorptivity measurement was $1$ mm, whereas it was $5$ mm in the emissivity measurements. A larger pinhole diameter is required in the emissivity measurements, in order to detect a sufficient signal from the nanomembranes that have lateral dimensions only $2$ mm $\times$ $2$ mm. For this reason, in Figs. \ref{fig2} (B), (E), and \ref{fig4} (B), the scale for the absorptivity and emissivity measurements, respectively, differ, and the emissivity lobes in Fig. \ref{fig4} (D) are broader than those of the absorptivity in Figs. \ref{fig2} (C), (F). In particular, this difference is attributed to the background thermal emission from the silicon frame of the membrane (see Fig. \ref{fig1} (A)) and the wafer, which are at the same temperature as the membrane, thereby also emitting IR light. This signal is obviously not present in the absorptivity measurements, when the sample is held at room temperature. Due to the $5$ mm pinhole used in the emissivity measurements, the signal from the silicon frame and wafer is considerable. In addition, we note that in contrast to the simultaneous measurements of emissivity and absorptivity, which were conducted within the FTIR microscope (Fig. \ref{fig3} (A)), the angle-dependent measurements of $\mathcal{E}$ and $\mathcal{A}$  were conducted separately, altering the sample position and alignment for every probed angle of observation and incidence, respectively.}

\par{In Fig. \ref{fig4} (C), we show the emissivity spectra for two observation angles, $\theta=36$ and $42$ degrees, and in Fig. \ref{fig4} (D) we show the corresponding polar plots of the thermal emissivity for the same wavelength as in Figs. \ref{fig2} (C), (F), i.e. for $\lambda=13.2$ $\mu$m (blue) as well as for $\lambda=13.7$ $\mu$m (black). The spectral width of the peaks shown in Fig. \ref{fig4} (C) are $1$ $\mu$m and $0.7$ $\mu$m, for $\theta=36$ and $42$ degrees, respectively, while the angular spread of the emission lobes shown in Fig. \ref{fig4} (D) are $12.9$ and $14.4$ degrees, respectively, for $\lambda=13.2$ $\mu$m and $\lambda=13.7$ $\mu$m, respectively. These spectral widths, corresponding to unpolarized light emitted from the sample are roughly equal to the sum of the spectral widths obtained when measuring absorptivity for TM and TE polarization separately, as shown in Figs. \ref{fig2} (C), (F), respectively. The increased angular spread with respect to the absorptivity measurements is associated with the aforementioned background parasitic emission from surrounding bodies. Despite this angular broadening, the highly directive character of the sample is preserved and clearly shown in thermal emissivity measurements. This angular spread is comparable with that reported via direct thermal emission measurements from diffraction gratings \cite{Greffet2002} as well as micro- and nano-structured surfaces \cite{Yu2023,Liu2019}.}

\section*{Discussion}

\par{We introduced an approach to generating highly directional and ultra-narrow-band infrared light through incandescence. In order to directly compare the proposed concept with the seminal work by Greffet \textit{et al.} \cite{Greffet2002} that serves as a hallmark in achieving directionality in thermal emission, in our experimental demonstration, the thermal emitter is also composed of SiC as in \cite{Greffet2002}. However, unlike \cite{Greffet2002} and relevant follow-up works \cite{Baranov2019, Tittl2015,Shen2016,Qu2020,Yu2023,Lu2021phononcoupling,Han:10,plasmonicmetagreffet}, the thermal source presented here does \textit{not} rely on the excitation of a surface-plasmon or phonon polariton, neither on diffraction, nor on the collective excitation of meta-atoms as in metasurface-based architectures. By contrast, it relies on wave interference and builds upon the well-established concept of a Salisbury screen \cite{salisbury_1952} that has been proposed in the 1950's. Via appropriate modifications to the characteristic dimensions of a Salisbury screen, thermal emission becomes highly spectrally coherent and directionally narrow for both linear polarizations.}

\par{With etching being the only synthesis step in the fabrication of the device, for the realization of an IR-wavelength-thick air gap, we propose a straightforward way to realize coherent incandescent sources at mid-IR wavelengths. The generated IR light is confined into a very narrow frequency range and angular range without the need of any high-resolution lithography. This creates a platform for ultra-efficient and simple-to-fabricate mid-IR thermal sources. The lateral size of the thermal source is on the order of millimeters, thus making this approach amenable to large-scale heat transfer applications and integrated mid-IR photonics at the wafer scale. The top layer, composed of a SiC nanomembrane in the present demonstration, can be replaced by another polar material with high-crystallinity, such as hexagonal boron nitride, $\mathrm{\alpha}$-Mo$\mathrm{O_3}$, and III-V materials like GaAs, InP and AlAs and others \cite{Sarkar2024, Caldwell2015}. Finally, the wavelength and angle of emission can be actively tuned by modifying the thickness of the air gap. This mechanism is available in micro electro-mechanical systems (MEMS) technology, thus enabling actively tunable coherent thermal emission sources. We hope that these findings can be a key factor in democratizing thermal infrared technologies, enabling cheaper IR spectroscopy and sensing, and improving integrated IR photonics functionalities.}

\section*{Acknowledgements}

\noindent{This work has been supported in part by la Caixa Foundation (ID 100010434),
the Spanish MICINN (PID2021-125441OA-I00, PID2020-112625GB-I00, and CEX2019-000910-S), the European Union (fellowship LCF/BQ/PI21/11830019 under the Marie Skłodowska-Curie Grant Agreement No. 847648), Generalitat de Catalunya (2021 SGR 01443), Fundació Cellex, and Fundació Mir-Puig.}

\bibliography{scifile}

\bibliographystyle{Science}


\end{document}